\documentclass[english,aps,preprint]{revtex4}
\usepackage[T1]{fontenc}
\usepackage[latin9]{inputenc}
\usepackage[active]{srcltx}
\usepackage{amsmath}
\usepackage{amssymb}
\usepackage{graphicx}
\usepackage{esint}

\makeatletter
\@ifundefined{textcolor}{}
{%
 \definecolor{BLACK}{gray}{0}
 \definecolor{WHITE}{gray}{1}
 \definecolor{RED}{rgb}{1,0,0}
 \definecolor{GREEN}{rgb}{0,1,0}
 \definecolor{BLUE}{rgb}{0,0,1}
 \definecolor{CYAN}{cmyk}{1,0,0,0}
 \definecolor{MAGENTA}{cmyk}{0,1,0,0}
 \definecolor{YELLOW}{cmyk}{0,0,1,0}
}

\makeatother

\usepackage{babel}
\begin{document}
\preprint{}
\title{Low-Frequency Characterization of Music Sounds\\
- Ultra-Bass Richness from the Sound Wave Beats - \\
}
\author{Masahiro Morikawa}
\email{hiro@phys.ocha.ac.jp}

\affiliation{Department of Physics, Ochanomizu University \\
2-1-1 Otsuka, Bunkyo, Tokyo 112-8610, Japan }
\begin{abstract}
Orchestra performance is full of sublime rich sounds. In particular,
the unison of violins sounds different from the solo violin. We try
to clarify this difference and similarity of unison and solo numerically
analyzing the beat of `violins` with timbre, vibrato, melody, and
resonance. Characteristic properties appear in the very low-frequency
part in the power spectrum of the wave amplitude squared. This ultra-buss
richness (UBR) can be a new characteristic of sound on top of the
well-known pitch, loudness, and timbre, although being inaudible directly.
We find this UBR is always characterized by a power-law at low-frequency
with the index around $-1$ and appears everywhere in music and thus
being universal. Furthermore, we explore this power-law property towards
much smaller frequency regions and suggest possible relation to the
1/f noise often found in music and many other fields in nature. 
\end{abstract}
\maketitle

\tableofcontents{}

\section{introduction}

The orchestra sound is very different from the solo-played sound.
In particular, the unison of the violin part sounds clearly different
from the solo violin. What is the difference and similarity between
the unison and the solo? The loudness of sound is apparently increased
in the orchestra, however, the violin unison in the orchestra has
mild tension and the sound quality is different from the solo sound.
The pitch and timbre of the violin seem to be the same for unison
and solo. We first study a possibility to introduce a new characteristic
of sound that may distinguish unison and solo. 

An important implication is the fact that there are few piano ensembles
(unison) except duets. One of the reasons would be the fact that the
piano cannot change its pitch while the violin and other instruments
can easily control its pitch. Therefore we speculate that the pitch
slightly different in the violin unison may be the key to solve this
problem. Actually, the multiple sound sources with slightly different
pitches with each other would cause the low-frequency beat. Larger
the number $N$ of sound sources, $_{N}C_{2}$ beat pattern appears
and the whole sound would become more complex and rich. We explore
this possibility: the low-frequency beat generated by the unison of
music instruments like a violin would characterize sound besides loudness,
pitch, and timbre. 

The superposition of the two simple sound sources, with slightly different
frequencies, can be described by trigonometric functions as $\sin(\omega t+\lambda t)+\sin(\omega t-\lambda t)$,
where $\omega\gg\lambda>0$. This is expressed as $2\cos(\lambda t)\sin(\omega t)$
and the factor $\cos(\lambda t)$ provides us with the low-frequency
beat. However, this beat cannot appear in a simple Fourier transformation
of the above formula; it trivially yields the frequency signal at
$\omega+\lambda$ and $\omega-\lambda$ only. However, if we take
the square of the sound wave amplitude, proportional to the intensity
of the wave, the beat does appear,

\begin{equation}
(\sin(\omega t+\lambda t)+\sin(\omega t-\lambda t))^{2}\xrightarrow[\text{Fourier tr. t\ensuremath{\rightarrow}k}]{}\sqrt{\frac{\pi}{2}}\delta(k\pm2\lambda)+...
\end{equation}
It often happens that the beat frequency $\lambda$ or $2\lambda$
is too small as an audible sound in the usual sense although it strongly
affects the musical impression as a modulation of the sound amplitude.
We call this characteristic the ultra-buss richness (UBR) property.
The above is the simple origin of the UBR of unison. 

Then what is the actual form of UBR in the case of many instruments?
In order to clarify the detail of the UBR, we need to take into account
the specific properties of the instrument and also the performance
style properties. There would be at least two such properties when
we study \textbf{unison}. 

a) One is the \textbf{timbre }of the sound, a superposition of higher
harmonics on top of the fiducial pitch. Each higher harmonics yields
a bunch of beats at ultra-low-frequency regions in the power spectrum
$S(\omega)$. Therefore the UBR reflects the power spectrum of higher
harmonics, which intrinsically depends on the musical instruments.
b) Another is the \textbf{vibrato}, a continuous modulation of low-frequency
for each note, a popular musical technique widely used nowadays. If
applied, the vibrato drastically increases the variety of frequency
differences in the ensemble even infinitely. This would make the UBR
richer. 

On the other hand, is UBR really specific to the unison? In order
to answer this question, we further consider the specific properties
of instruments and performance style properties in the case of solo.
There would be at least two such properties when we study \textbf{solo}. 

a) One is the \textbf{melody flow}. We would like to consider a sequence
of notes, a melody, and the possibility of overall UBR, even if each
note of solo instrument may not possess UBR. It may be particularly
interesting, for example, to try to make each adjacent notes slightly
overlap with each other in the melody to observe any UBR. Since this
is a typical case of musical instruments, such as multi-string violin,
this example is also practical. b) Another is the \textbf{resonance}
of a single instrument. Even if a single string posses a single fiducial
frequency $\omega$, as well as overtones, the resonating box attached
to the string would have some finite range of resonance frequency
around $\omega$. If these continuous resonance modes are excited
and cause beats with each other, then we can expect UBR even for a
single sound source instrument. 

The emphasis in this paper is the sound property UBR, the low-frequency
characteristic in the power spectrum. This makes UBR unique distinguished
from the other three characteristics of sound: loudness, pitch, and
timbre, all of these characterize the sound at the frequency $\omega$
or higher in the power spectrum. Further, our interest naturally continues
toward much lower frequency regions. 

In this context, it is widely known that 1/f or pink noise appears
very often in the ultra low-frequency regions and is characterized
by the universal power-law in the power spectrum: $S(\omega)\propto\omega^{\beta}$
with $\beta\approx-1\sim-1.5$\cite{Milotti2002}. In particular,
this 1/f fluctuation is reported also to appear in music especially
in classical music\cite{Musha1981}. 

The case $\beta=-1$ is particularly interesting since the integrated
fluctuation $\intop_{0}^{\infty}S\left(\omega\right)d\omega$ would
diverge in both low and high frequency ends. Similar case appears
in the early Universe in the spatial domain. The density fluctuations
of (dark) matter generated from quantum mechanics have divergent integrated
fluctuation in both the small and large scale ends (Zel'dovich spectrum)\cite{Dodelson2020}.
Since our sound beat creates UBR, we may be able to expect some relation
between UBR and the 1/f noise. We would like to explore to some extent
this relation. 

This paper is composed as follows. Section 2 describes how UBR appears
in the unison of the sound sources which have timbre and vibrato.
Section 3 describes the possibility of the UBR in the solo sound for
the cases of melody and resonance. In section 4, we analyze real music
in the same way as above and try to verify our point of view. In section
5, we explore the possibility that the wave beat yields the 1/f fluctuations
in general. The last section 6 summarizes our study showing possible
extensions of the present calculations. 

\section{UBR from unison }

\subsection{Unison with timbre (tone color) }

We first study the unison of multiple sound sources which have timbre
or overtones. The timbre is made from the superposition of multiple
harmonics. In the case of the violin, the wave amplitude of the $n$-th
harmonics can be approximated to be proportional to $n^{\beta}$ with
$\beta\approx-0.7$ \cite{yokoyama2016}, although actually, some
overtones deviate from this power-law and the power index itself varies
from $-0.5$ to $-1.5$ in numerous literature. 

We simply superpose trigonometric overtone wave with the weight $n^{\beta}$,where
$\beta=-0.7$ to form a `violin` sound $v_{0}\left(\omega\right)$, 

\begin{equation}
v_{0}\left(\omega,t\right)=\sum_{m=1}^{M}n^{\beta}\sin\left(2\pi m\omega t+\eta\right),\label{eq:v0}
\end{equation}
where $\omega$ is the fiducial frequency, $\eta\in[-\pi,\pi]$ is
a random phase, and we superpose up to $M$-th overtone. Then, $N$
times superposing them with random modulation $\xi$ (within some
range), 
\begin{equation}
v\left(t\right)=\sum_{\mathrm{random}\xi,\eta}^{N}v_{0}\left(\omega+\xi\right),\label{eq:v}
\end{equation}
we obtain a unison sound of `violins` with timbre. 

Now we analyze this sound in the Fourier power spectrum. Analytic
Fourier transformation is possible but the results are a collection
of cumbersome terms and are intractable. Thus we use, in this paper,
discrete Fourier transformation with sufficient sampling points. As
was explained in the previous section, a simple Fourier transformation
of $v\left(t\right)$ trivially yields no signal in the low-frequency
regions. Thus we always take discrete Fourier transformation for the
square of the data $v\left(t\right)^{2}$. There is another indicator
that also characterizes the power distribution: zero-crossing which
simply counts the zero of the data. The power spectrum of this zero-crossing
shows similar behavior as $v\left(t\right)^{2}$ although we do not
discuss this zero-crossing indicator in this paper. 

\begin{figure}
\includegraphics[height=15cm]{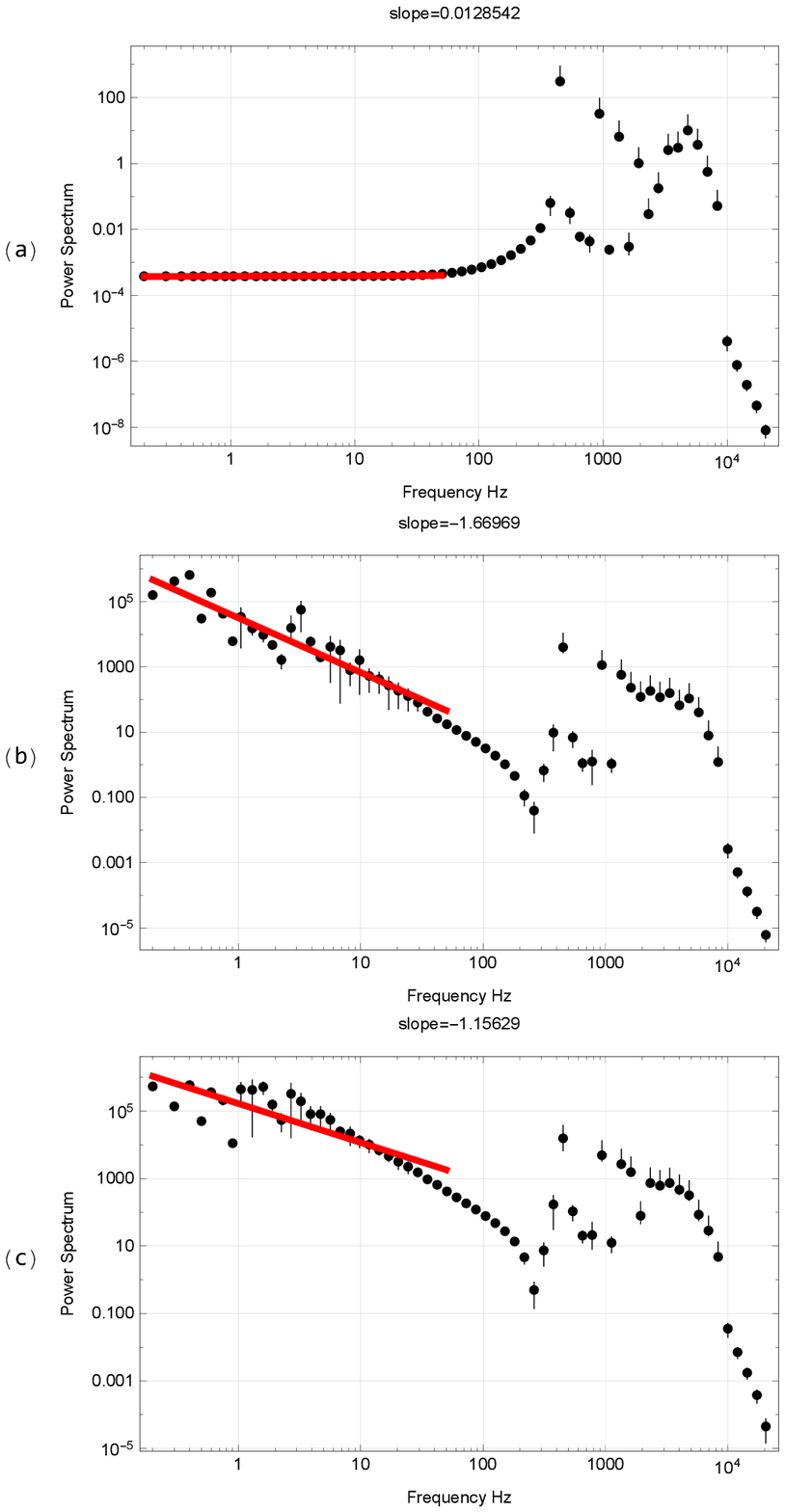}

\caption{Power spectrum (PS) of simple `violin` with timbre. \protect \\
a) PS of violin solo ($N=1$). It never yield low-frequency signal
enhancement UBR. \protect \\
b) same but violin quintet ($N=5$). UBR starts to appear in the form
of power-law with negative index about $-1.7.$\protect \\
c) same but 10 violins ($N=1$0). UBR develops in the form of power-law
but becomes shallower with the index about $-1.2$. \protect \\
The parameters used in the numerical calculation of Eq.(\ref{eq:v})
are, $\omega=440,-3<\xi(\mathrm{random})<3,\beta=-0.7,\tau=10,M=30$.
Each violin has different pitch within 0.7\% from the fiducial $440$Hz.
The data time series is prepared with the sampling rate $44100$,
i.e. $44100$ times sampling within a second before the discrete Fourier
transformation of the data $v\left(t\right)^{2}$. The data is generated
by Wolfram Mathematica12. \protect \\
\label{violin timbre}}
\end{figure}

A single `violin` with timbre never shows any signal in ultra bass
(UB) region in the power spectrum (as well as zero-crossing data).
This is shown in Fig. \ref{violin timbre} (a), where we take the
parameters $\omega=440,-3<\xi(\mathrm{random})<3,\beta=-0.7,\tau=10,M=30,N=1$,
where $\tau$ is the time duration, in second, of the sound data.

On the other hand, when we superpose the sounds of 5 `violins`, the
strong signal appears in the low-frequency regions, as is shown in
Fig. \ref{violin timbre} (b). The UBR starts to appear. This UBR
is well characterized by the power-law with the power index of about
$-1.7$ within the 3 decades, while the timbre range, for $M=30$,
is about 1.5 decades. This long-decade power-law of UBR comes from
the variety of all possible beat pairs. Although the modulation range
is fixed within $0.7\%$ in the present case, a real player in an
orchestra tries to tune to the fiducial pitch as much as possible.
Then the variety of beat continuously increase and therefore the UBR
will extend much lower frequency regions. 

Further, the superposition of 10 `violins` shows the stronger signal
of UBR as is shown in Fig. \ref{violin timbre} (c). We have played
around various parameters including those in Fig.\ref{violin timbre},
and have found that the power index $\gamma$ generally increases
(\textit{i.e.} becomes shallower) up to about $-1$, for a larger
number of violins. 

\subsection{Unison with vibrato }

We now study unison with vibrato, a popular performance style, being
a pitch fluctuation technique by the player. A uniform vibrato of
frequency $\theta$ with amplitude $b$ on top of the fiducial frequency
$\omega$ can be expressed as 

\begin{equation}
\begin{aligned}\omega_{vib}\left(\omega,\theta,b,t\right)= & \int_{0}^{t}2\pi\left(\omega+b\left(\sin(2\pi\theta t'+\eta)\right)\right)\,dt'\\
= & \frac{2b\sin(\pi\theta t)\sin(\pi\theta t+\eta)}{\theta}+2\pi\omega t
\end{aligned}
\label{eq:omegavib}
\end{equation}
where $\eta\in[-\pi,\pi]$ is a random phase. Then, $N-$times superposing
them with random phases $\xi$, 

\begin{equation}
v\left(t\right)=\sum_{\mathrm{random}\xi,\eta}^{N}\sin\left(\omega_{vib}\left(\omega+\xi,\theta,b,t\right)\right)\label{eq:vomegavib}
\end{equation}
we obtain a unison sound of `violins` with vibrato. 

\begin{figure}
\includegraphics[height=15cm]{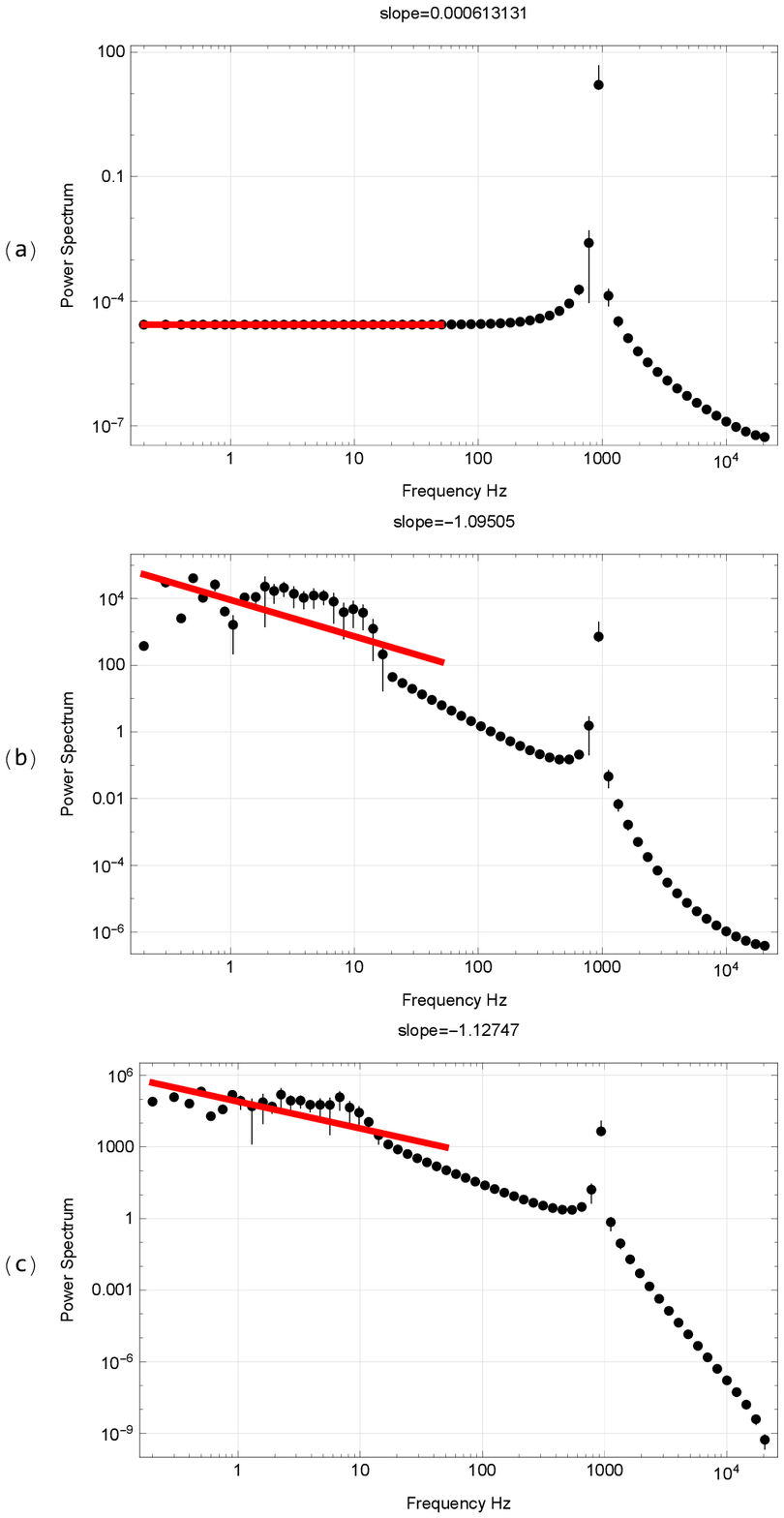}

\caption{Power spectrum (PS) of simple `violin` with vibrato. The data time
series is prepared with the sampling rate 44100, i.e. 44100 times
sampling within a second before the discrete Fourier transformation
for PS. Each violin has different pitch within 3\% from the original
440Hz. \protect \\
a) PS of a single violin ($N=1$). It never yield UBR alone. \protect \\
b) same but violin quintet ($N=5$).UBR starts to appear in the form
of power-law with negative index about $-1.1.$\protect \\
c) same but 10 violins($N=1$0). UBR develops in the form of power-law
with negative index about $-1.1$.\protect \\
UBR is characterized by the power-law with power index about $-1.1$
for unison. The parameters are $\omega=440,b=2+(-1<\mathrm{random<1}),-1<\theta(\mathrm{random})<10,$$-6<\xi(\mathrm{random})<6,\tau=10,N=1,5,10$.
\label{violin vibrato}}
\end{figure}

Now we analyze this sound in the Fourier power spectrum as before.
A single `violin` with vibrato never shows any signal in ultra bass
(UB) region in the power spectrum as previously. This is shown in
Fig. \ref{violin vibrato} (a), where we take the parameters $\omega=440,b=2+(-1<\mathrm{random<1}),-1<\theta(\mathrm{random})<1,-6<\xi(\mathrm{random})<6,\tau=10,N=1$. 

On the other hand, if we add the sounds of 5 violins (b), or of 10
violins (c), UBR evolves in the form of a power-law with characteristic
power indexes about $-1.1$ as in Fig.\ref{violin vibrato}(b)(c). 

\subsection{Unison with timbre and vibrato }

\begin{figure}
\includegraphics[height=15cm]{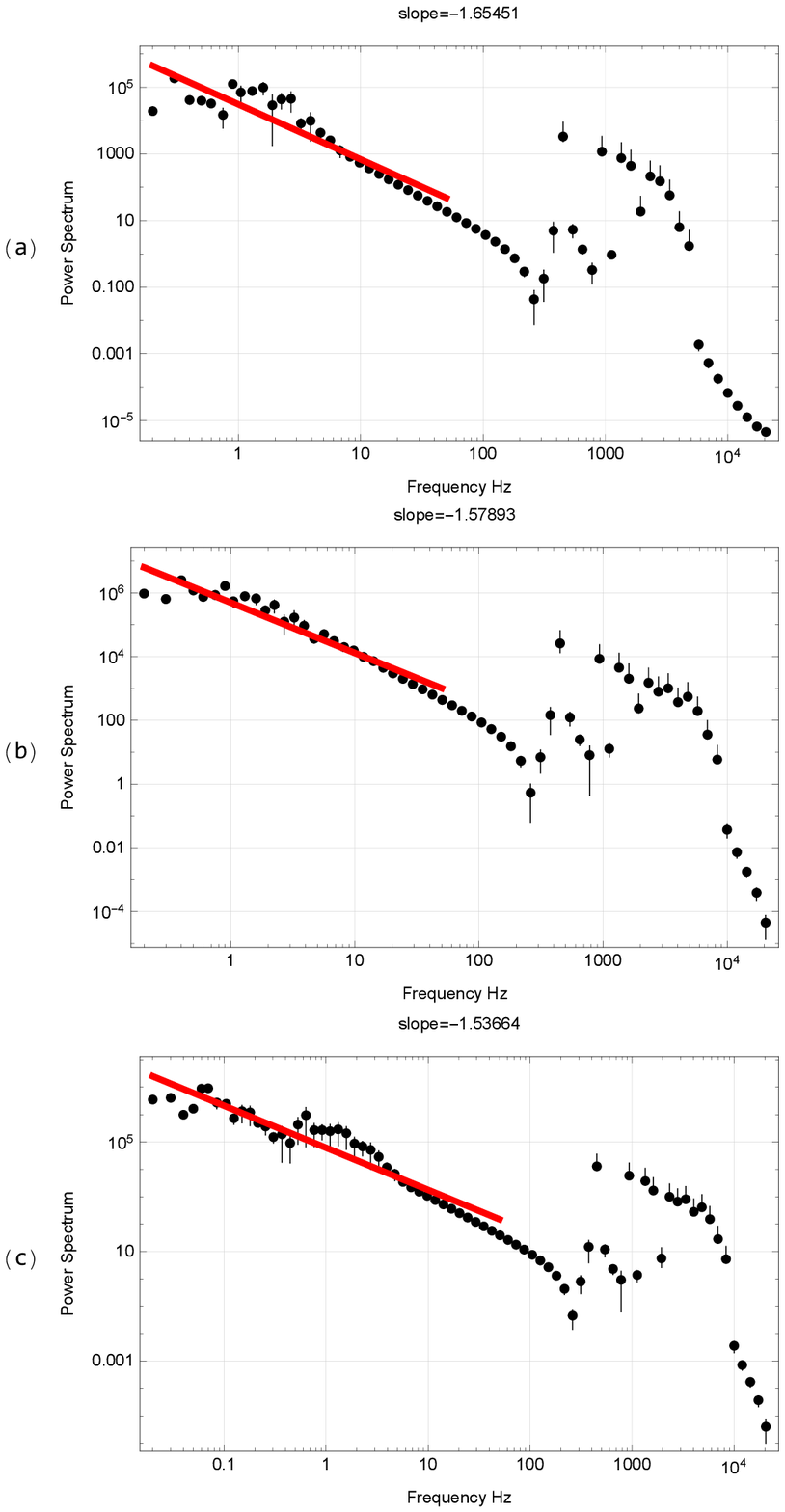}\caption{The power spectrum of `violin` sound data squared with timbre and
vibrato. \protect \\
(a) 5 violins superposed with the parameters:\protect \\
 $\omega=440,$$-1<\xi(\mathrm{random})<1,\beta=-0.7,\tau=10,M=5,N=5$.
\protect \\
(b)10 violins, of more overtones, superposed with the parameters:\protect \\
 $\omega=440,$$-1<\xi(\mathrm{random})<1,\beta=-0.7,\tau=10,M=10,N=10$.
\protect \\
(c)long time scale data for 10 violin sound superposed, with the parameters:\protect \\
 $\omega=440,$$-0.1<\xi(\mathrm{random})<0.1,\beta=-0.7,\tau=100,M=10,N=10$.
\protect \\
We set the pitch fluctuation smaller than that for shorter time scale
$\tau$. This makes the dispersion of the data small. The power index,
which characterize UBR, is about $-1.7\thicksim-1.5$. \label{violin timbre vibrato}}
\end{figure}

We now consider unison with timbre and vibrato together for more realistic
violin performance. Combining Eq.(\ref{eq:omegavib}) and Eq.(\ref{eq:v}),
and randomly superposing them to make unison, we have the sound wave, 

\begin{equation}
v=\sum_{\mathrm{random}\xi,\eta}^{N}v_{0}\left(\omega_{vib}\left(\omega+\xi,\theta,b,t\right)+\eta\right).
\end{equation}
The square of this data yields the PS as in Fig.\ref{violin timbre vibrato}.
In all the panels (a)(b)(c), UBR is clearly observed in the form of
power-law with indexes $-1.7\sim-1.5.$ Further, the dispersion of
low-frequency data seems to be smaller for larger number of violins
as in panels (a)and (b). 

We tried to take data longer in time in Fig.\ref{violin timbre vibrato}(c).
UBR seems to continue toward much smaller frequencies although we
set the frequency fluctuation small to reduce the dispersion of the
data. It may happen that the violinists in an orchestra may try to
do the same; to tune more for longer unison. 

In conclusion of this section, unison always seems to yield UBR and
characterized by a power-law with index $-1.1\sim-1.7$. 

\section{UBR from Solo \label{sec:UBR-from-Solo}}

We have always found UBR in unison(Fig.\ref{violin timbre}(b)(c),
Fig.\ref{violin vibrato}(b)(c)) but not in solo(Fig.\ref{violin timbre}(a),
Fig.\ref{violin vibrato}(a)), within our study so far. Then, can
we finally conclude that no UBR is observed in solo? Is UBR really
limited to unison? Or, can UBR be universal to some extent? To answer
these questions, in this section, we study solo cases considering
more realistic conditions than the previous simplest settings. 

\subsection{melody of solo violin \label{subsec:melody-of-solo}}

A solo `violin` so far we artificially constructed did not show UBR
at all(panels (a) of Figs.\ref{violin timbre},\ref{violin vibrato}).
However, music sounds always appear in a melody or a sequence of notes.
Quoting a short melody from J.S.Bach Partita No.2 in D Minor, BWV
1,  we first make a sequence of sounds $re-mi-fa-re-re-do-\natural si-la-$,$\sharp so-la-\natural si-la-so-\sharp fa-mi-re$. 

\begin{figure}
\includegraphics[height=15cm]{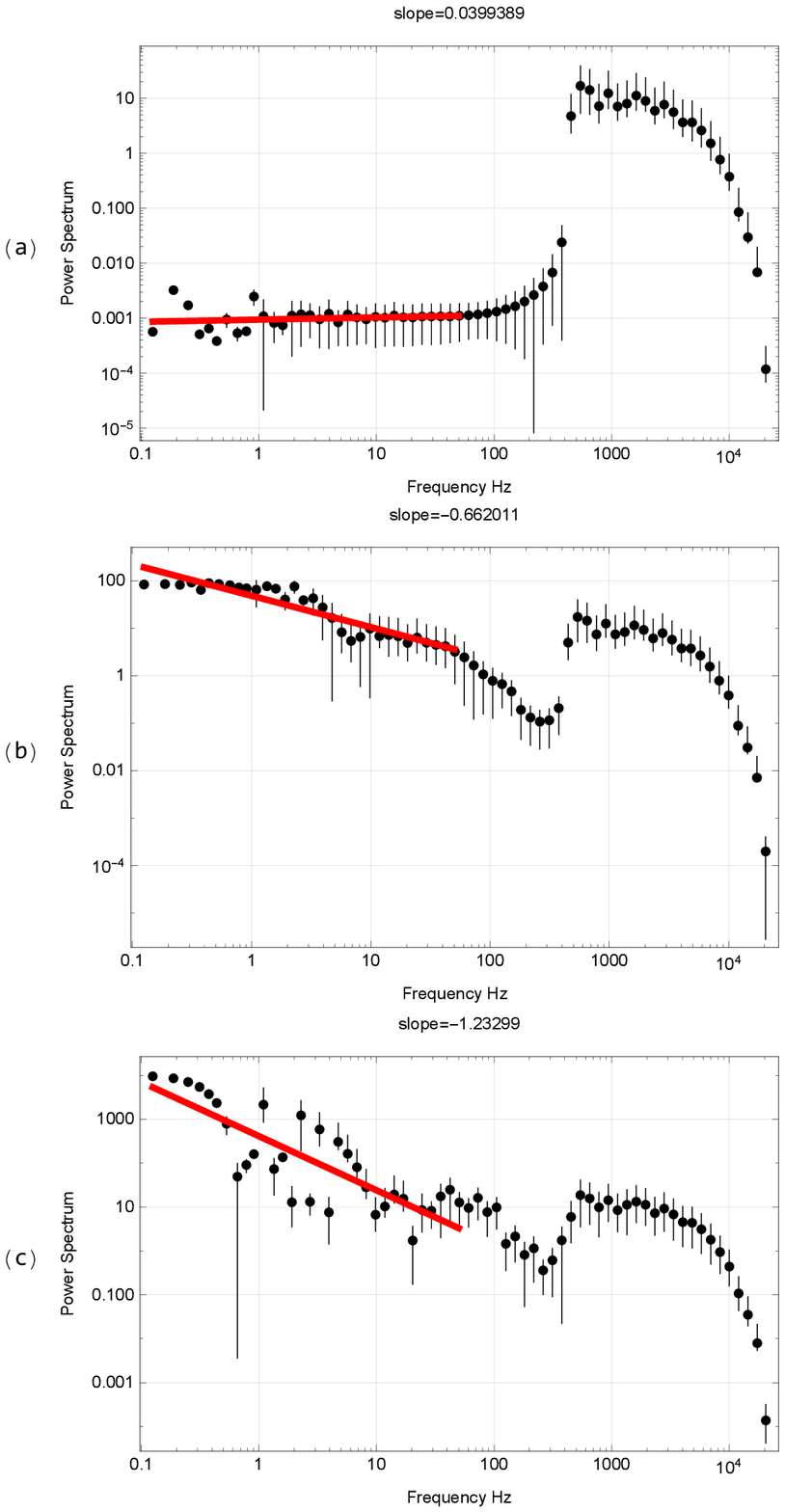} \\
\caption{The PS of the melody $re-mi-fa-re-re-do-\natural si-la-$,$\sharp so-la-\natural si-la-so-\sharp fa-mi-re$.
generated in the equal temperament. \protect \\
(a) PS of a simple melody without overlap. UBR does not appear. \protect \\
(b) Same as (a) but each note has 1\% overlap with the adjacent notes.
UBR does appear and the power index is $-0.7$. \protect \\
(c) Same as (b) but with 10\% overlap. UBR is enhanced and the power
index is $-1.2$. \protect \\
A note of $la$ in the melody is generated by the `violin` note with
timbre Eq.(\ref{eq:v}) and the parameters are, $\omega=440,-3<\xi(\mathrm{random})<3,\beta=-0.7,\tau=1,M=10,N=1$.
Based on this, we made the other notes in equal temperament, \textit{i.e.}
multiplying necessary powers of $2^{1/12}$ to the fiducial pitch.
\label{melody}}
\end{figure}

We first simply adjoined each note to form a melody data and the square
of it is analyzed by the discrete Fourier Transformation to get the
power spectrum. As expected, this does not show UBR at all (Fig. \ref{melody}
(a)). 

However, the actual instrument performance of melody may not make
each note finish exactly before the beginning of the next note. Therefore,
we simply make tiny overlap adjacent notes with each other. If we
take this overlap as 1\% of each note segment length, we have the
PS in Fig.\ref{melody} (b). There drastically appears clear UBR only
by 1\% segment overlap. 

If we further superpose each note to 10\%, we have the PS in Fig.\ref{melody}
(c). The UBR is enhanced than case (b). Further, in these cases, UBR
is characterized by the power-law with the power index $\gamma=-0.7\thicksim-1.2$. 

The reason that a melody as a simple sequence of notes does not show
UBR will be the notes are statistically independent with each other.
If they follow some statistical rule, then UBR may appear. We will
examine this possibility in section \ref{sec:Beyond-music}. The key
reason that a melody with each adjacent segment is overlapped a bit
does show UBR will be the existence of any statistical correlation
among very small overlapped notes. There is a possibility that an
each element seems to be random but the whole sound data may follow
any statistical rule. In this case, there may appear long period correlations
that characterize the very low-frequency regions in PS. This point
will be further studied in section \ref{sec:Beyond-music}. 

An important caution should be made here. We have treated the melody
composed from the notes of equal strength in our analysis and have
concentrated on the UBR from the sound beat. However, real music is
rich and has its own expression, crescendo/decrescendo or swinging
tempo, as well as the hierarchical structures as a composed piece,
from motive to the entire symphony. This musical structure itself
may yield UBR in particular at very-low-frequency regions\cite{Musha1981}.
This latter UBR should be separated from our present analysis. 

\subsection{resonance of solo violin \label{subsec:resonance-of-solo}}

A string of a violin is not isolated but is attached to a wooden box
which makes resonance and emits many notes with slightly different
frequency from the string pitch. We consider the possibility of unison
made from the resonance. By solving a forced harmonic oscillator,
a simple resonance would yield the following sound, 
\begin{equation}
v\left(t\right)=\sum_{random\xi}^{N}\left(\frac{\lambda(\sin(2\pi\text{\ensuremath{\omega}}t))}{\left(\omega+\xi\right)^{2}-\omega^{2}}+\sin(2\pi t\text{\ensuremath{\left(\omega+\xi\right)}})\right),\label{eq:resonance}
\end{equation}
where $\omega$ is the forcing frequency, corresponding to the string
frequency, and $\lambda$ is the coupling to yield resonance. 

If we include timbre, then the sound is given by
\begin{equation}
v\left(t\right)=\sum_{m}^{M}m^{\beta}\sum_{random\xi}^{\text{N}}\left(\frac{\lambda\sin(2\pi\text{\ensuremath{m\omega}}t)}{(m\left(\omega+\xi\right))^{2}-(m\omega)^{2}}+\sin(2\pi tm\left(\omega+\xi\right))\right).\label{eq:resonance +timbre}
\end{equation}
If the resonance is further dissipative, we replace the factor $(m\left(\omega+\xi\right))^{2}-(m\omega)^{2}$
and obtain 
\begin{equation}
v\left(t\right)=\sum_{m}^{\text{M}}m^{\beta}\sum_{random\xi}^{\text{N}}\left(\frac{\lambda\sin(2\pi m\text{\ensuremath{\omega}}t)}{\sqrt{\left((m\left(\omega+\xi\right))^{2}-(m\omega)^{2}\right)^{2}+4\mu^{2}(m\omega)^{2}}}+\sin(2\pi tm\left(\omega+\xi\right))\right),\label{eq:resonance + timbre+dissipation}
\end{equation}
where $\mu$ is the dissipation constant. 

\begin{figure}
\includegraphics[height=15cm]{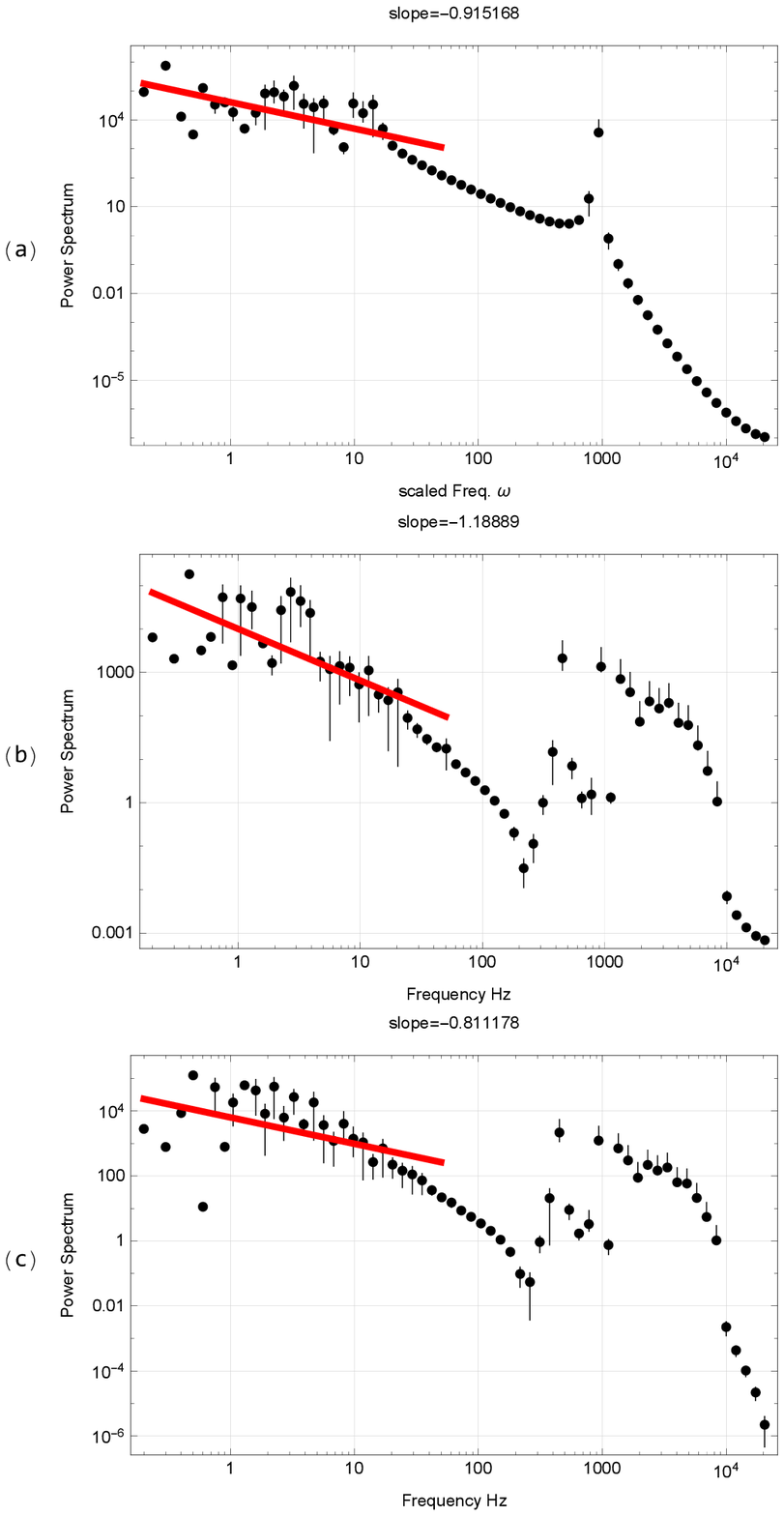}\caption{A single `violin` with resonance can yield UBR which is characterized
by the approximate power-law with index $-1.2\sim-0.8$. \protect \\
(a)PS of a simple case along with Eq.\ref{eq:resonance} with parameters
$\omega=440,\lambda=10,\xi\in[-10,10],N=10,\tau=10$ . \protect \\
(b)PS of a case also with timbre along with Eq.\ref{eq:resonance +timbre}
with parameters $\omega=440,\lambda=10,\xi\in[-3,3],N=10,M=5,\tau=10,\beta=-0.7$.
 \protect \\
(c)PS of a case also with timbre and dissipation along with Eq.\ref{eq:resonance + timbre+dissipation}
with parameters $\omega=440,\lambda=10,\xi\in[-3,3],N=10,M=5,\tau=10,\beta=-0.7,\mu=10$
.  \label{resonance}}
\end{figure}

These sound data Eqs.(\ref{eq:resonance}, \ref{eq:resonance +timbre},
\ref{eq:resonance + timbre+dissipation}) yield PS in Fig.\ref{resonance}
in order. We found that even a solo `violin` with resonance can yield
UBR, and again this UBR is characterized by the approximate power-law
with index $-1.2\sim-0.8$. 

Although we have examined only a few cases for solo instruments in
this paper, there will be many cases that solo instruments that may
yield UBR. For example, reflection from a wall in a hall can easily
form a superposition of sounds, resonance with the hall, and the other
instruments can naturally form a superposition of sounds. Therefore,
music performance is thought to be full of UBR. 

Concluding this section, even a single `violin` can yield UBR and
it is again characterized by a power-law with an index of about $-1$.

\section{Some verification in real music}

The study so far is based on the artificial sound generated by computers.
We briefly examine UBR in real music performances in this section. 

In the case of \textbf{orchestra unison}, it is difficult to find
a pure unison of a single pitch. Therefore, instead, we randomly clip
a segment from the full performance for the analysis. The data is
from you-tube provided in the wav-format and we used a single channel
from stereo-recorded data. The sampling rate is 44100 Hz for all the
data, sufficient to analyze the low-frequency regions. 

\begin{figure}
\includegraphics[height=15cm]{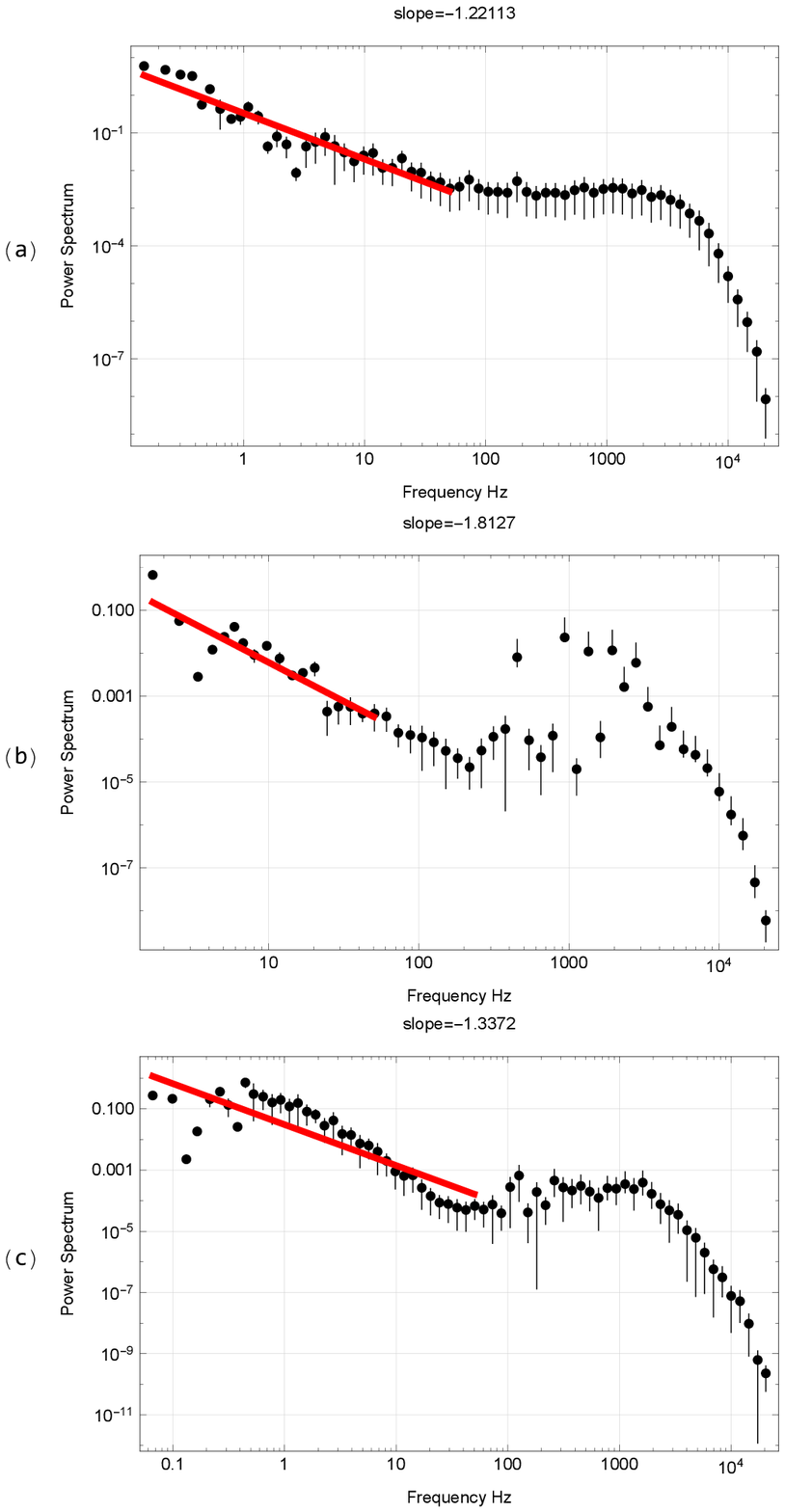}\caption{PS of some real instrument performances. The data are from you-tube
in wav-format, and we used a single channel data of sampling rate
44100 Hz. We have randomly clipped a part from the full performance
for the analysis. \protect \\
(a)PS of the sound of orchestra performance. A 13-second clip from
\textquotedbl Tchaikovsky Serenade for Strings Ozawa Saitoh-Kinen
Orchestra\textquotedbl\cite{Ozawa1992}. \protect \\
(b)PS of solo violin sound 1.2-second clip of a single tone\cite{Kuboki2014}.
\protect \\
(c)PS of solo violin sound 30-second clip from J. S. Bach Partita
for Solo Violin No. 3\cite{Perlman2012}. \label{PS real performance}}
\end{figure}

Results are in Fig.\ref{PS real performance} which shows PS obtained
from the sound data squared applying the discrete Fourier transformation.
The top panel (a) shows the PS of a 13-second clip of orchestra performance.
We choose Tchaikovsky Serenade for Strings conducted by S. Ozawa played
by the Saitoh Kinen Orchestra from YouTube \cite{Ozawa1992}. This
shows clear UBR in the form of power-law with the index $-1.2$. However,
the actual orchestra performance, including the concert hall and the
location of the audience, is very complex and therefore cannot directly
be compared with our analysis. 

In the case of \textbf{solo violin, }we first clip a pure segment
of sound with a single pitch. The power spectrum is in Fig.\ref{PS real performance}
(b). The data is a short clip of 1.2 seconds at the scene of almost
constant pitch\cite{Kuboki2014}. Clearly, we observe UBR with power-law
but with a slightly steep power index $\gamma=-1.8$. The other clips
also show a similar tendency with the power index $\gamma=1.5-2.1$.
This should be compared with section III.B Fig.\ref{resonance}, the
resonance of solo violin. In particular, in the corresponding region
of 1 to hundred Hz, the slope becomes steeper and the index $\gamma$
is about $-2$. in Fig.\ref{resonance}. This tendency should be compared
with the above slightly steep power indexes. 

In the case of \textbf{solo violin with melody, }we choose J. S. Bach
Partita for Solo Violin No.3 - from YouTube \cite{Perlman2012}. The
power spectrum is in Fig.\ref{PS real performance}(c) which shows
UBR. This should be compared with the arguments in III.A., in particular,
Fig. \ref{melody} (c). 

Although we also tried to use the sound data itself, not squared,
no UBR was found. We further tried to use the zero-point method as
explained before and obtained a similar behavior as in Fig.\ref{PS real performance}.
These properties are the same as all our analyses in the previous
sections. 

\section{Beyond music \label{sec:Beyond-music}}

We have so far studied the beat of multiple sound waves at low-frequency
regions and found the ultra-bass richness (UBR) in the power-law form
in PS. The essence of UBR is the wave beat and this simple principle
makes the UBR universal. The UBR for the sound waves is characterized
by a power-law with its index $\gamma=-1.5\thicksim-1$. This power-law
may extend further much lower frequency regions as we have seen in
Fig. \ref{violin timbre vibrato}(c). 

These properties remind us of 1/f fluctuations or pink noise characterized
by the low-frequency power-law with index $-1.5\thicksim-1,$ distinct
from the white noise (index $0$) or Brownian noise (index $-2$).
This pink noise seems to appear everywhere in nature and to have many
kinds of origins\cite{Milotti2002}. 

We pick up some typical examples of 1/f fluctuation: vacuum tube,
semiconductor, human heart, squid giant axon, brain MEG and EEG\cite{Johnson1925,Linkenkaer2001,Musha1981,Novikov1997},...
The first two of these directly measure the electric current. The
others are related to the nerve of living creatures and the signal
passing through the nerves should be in the form of an electric current.
Therefore, all of these seem to be related to the fluctuations of
the electric current. Further, the electric current density $j^{\mu}=e\overline{\psi}\gamma^{\mu}\psi$
is the square of the electron wave functions $\psi$ in the fundamental
level. Therefore, it will be natural to speculate that the electron
waves in the object make beats to yield UBR. In particular, the beats
are produced by the electromagnetic scattering process with photon,
which has common dispersion relation with sound\cite{Handel1980}
at low-frequency, and therefore infrared-divergent noise arises. However,
these electron waves, at least, cannot be the quantum mechanical wave
functions\cite{Nieuwenhuizen1987,Kiss1986}. 

Here in this section, we explore a general possibility that UBR caused
by infrared-divergent noise, reserving the quantum mechanical origin. 

\begin{figure}
\includegraphics[height=15cm]{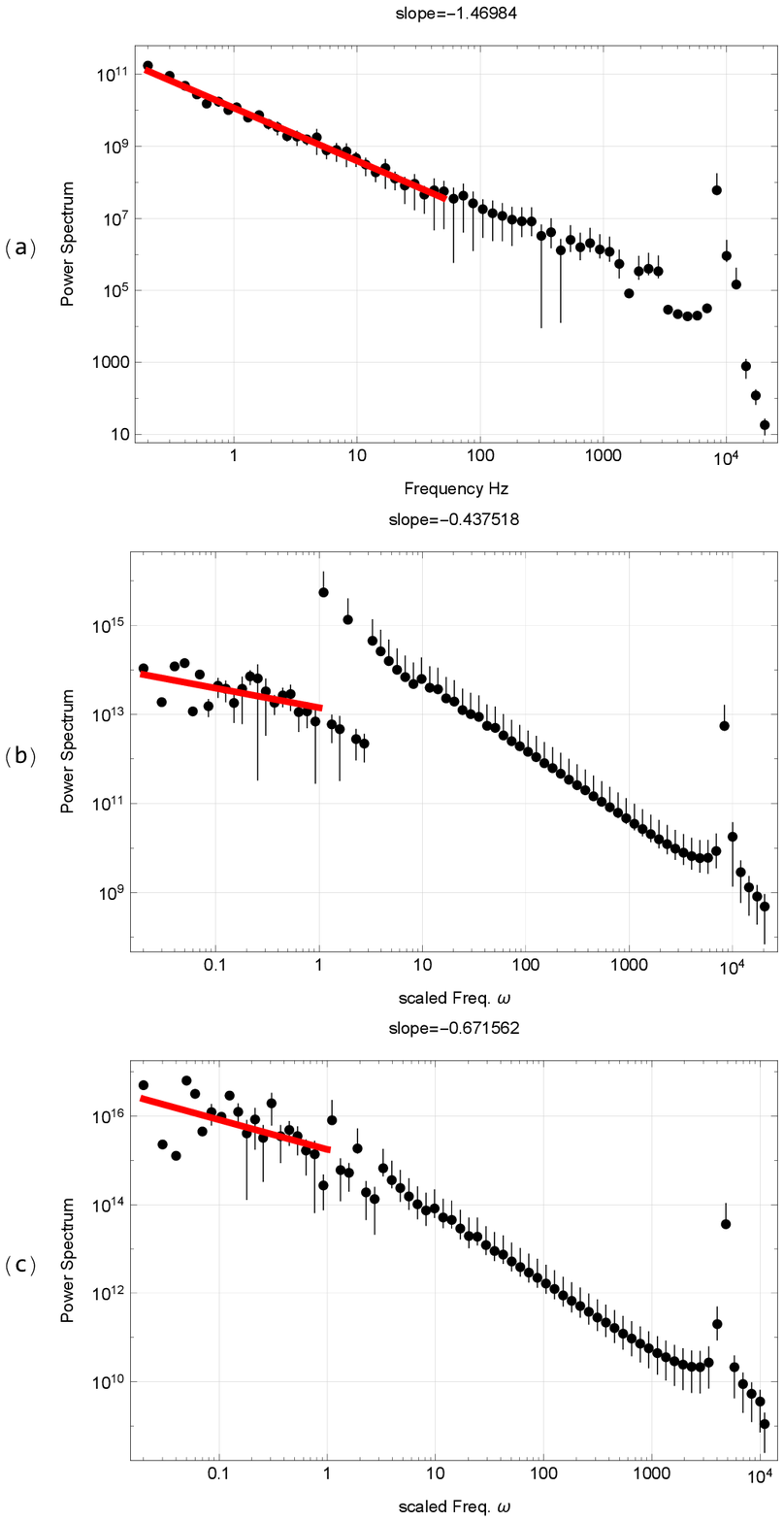}\caption{PS of the data generated by Eq.\ref{eq:irdiv}. This data represents
the beat or interference among many scatted wave modulated by the
noise which has infrared-divergent distributions. \protect \\
(a) PS for a single data of time duration $\tau=10$ second. The parameters
are $\omega=4400,-3000<\xi(\mathrm{random})<3000,\tau=10,N=300,\epsilon=10^{-5}$.
It shows a clear power-law with index $-1.5$. \protect \\
(b) PS of data which is a sequence of independent 100 segments each
of it is generated by Eq. \ref{eq:irdiv} with time duration $\tau=1$
second. There seems to appear another power with shallower slope in
the low-frequency region below 1 Hz which goes beyond the causality
limit 1 second. The parameters are $\omega=4400,-12400<\kappa(IR-div)<12400,\tau=1,N=4096,\epsilon=10^{-5}$.
\protect \\
(c) PS generated by the data of (b) but adjacent segments are 50\%
superposed with each other. \label{irdiv}}
\end{figure}

The infrared-divergent distribution function, with cutoff $\epsilon,$
is $\left(\omega+\epsilon\right)^{-1}$ and the inverse function of
it's cumulative distribution function $\epsilon\left(e^{x}-1\right)$
generates such infrared-divergent random field $\kappa$ from the
uniform random field. The N-superposed data is thus given by 
\begin{equation}
v\left(t\right)=\sum_{random\kappa,\eta}^{N}\sin\left(2\pi t\text{\ensuremath{\left(\omega+\kappa\right)}+\ensuremath{\eta}}\right).\label{eq:irdiv}
\end{equation}
The power spectrum of this data is shown in Fig.\ref{irdiv}, in which
there appear UBR that is expressed as the power-law. The top panel
(a) is the PS of the square of full data Eq.(\ref{eq:irdiv}), and
shows a clear power-law with index $-1.5$. 

When we consider the finite size of the sound source $\lambda_{0}$,
it seems natural to raise an objection for the above appearance of
UBR below the critical frequency $\omega_{0}\equiv2\pi/\lambda_{0}$
which corresponds to the scale $\lambda_{0}$. However, it should
be noted that the noise $\kappa$ is generated by an infrared-divergent
distribution function which has no fixed finite mean nor standard
deviations for a finite-size sample. Therefore, even if we construct
the whole data directly connecting many independent segments of size
$\lambda_{0}$, the whole data will show any statistical property
reflecting the infrared diverging noise toward the lower scale beyond
the critical frequency $\omega_{0}$. Thus we expect some kind of
UBR, which may not be the original type, in such data composed from
many independent segments. A typical example is shown in Fig.\ref{irdiv}(b)
where another type of UBR seems to appear separately in the lowest
frequency regions. 

This low-frequency UBR becomes more prominent when we allow an overlap
of each adjacent segment in the data as we did before in the melody
section \ref{subsec:melody-of-solo}. A typical example is shown in
Fig.\ref{irdiv}(c), where the PS is shown for the same data of (b)
but allowing 50\% overlap for each segment. 

Thus, we confirm that the size of the sound source $\lambda_{0}$
will not be a limitation for the appearance of UBR toward the lower-frequency
regions below $\omega_{0}$. This may be a challenge to one of the
physical objections \cite{Nieuwenhuizen1987,Kiss1986} for quantum
1/f noise\cite{Handel1980}. 

\section{Summary and Future prospects}

We started our study with a naive question: how unison is different
from solo. At first, we expected the unison or a superposition of
slightly different many pitches would make a beat and naturally yields
a clear signal in the low-frequency regions UBR, while solo does not
have this property. However, it turned out that even solo can yield
UBR through a superposition of slightly overlapped notes (melody)
or resonance, though the signal is not so strong as the unison case.
We found UBR is often characterized by a power-law with index $-1.5\thicksim-1$.
Then we have briefly verified these general arguments by using actual
music sound sources. 

Then we examined a possible link from our study on musical sound to
the general 1/f noise, expecting that the 1/f noise is generated by
wave beats. We have demonstrated this possibility within limited examples.
Introducing an infrared-divergent noise, we studied how it can yield
UBR. We found clear UBR in the data even if it is decomposed into
independent segments, in particular when we allow superposition among
adjacent segments. 

The UBR we studied in this paper may be the fourth element to characterize
the sound as well as the traditional three elements; sound loudness,
pitch, and timbre. All of these traditional elements characterize
sound, in the PS, at the original frequency (pitch, loudness) and
higher (timbre) while UBR at all lower frequency regions than the
original. If allowed to be personal and emotional, UBR can be described
as a profound dignified sound with comfortable mild tension. Our perspective
is that multiple sound beats awaken these emotions. 

Our study of UBR has just started from this paper and there remain
many subjects necessary for extensions and versification. Such issues
are in order below. 
\begin{enumerate}
\item Investigations for various \textbf{musical instruments} are necessary.
For timbre, we made a violin-like sound by superposing $n-$th overtones
with specific weight $n^{\beta}$ with $\beta=-0.7$. However, the
value of $\beta$ is different for other instruments. UBR power index
$\gamma$ depends on $\beta$. If the flute has larger $\beta$, then
the same for $\gamma$ and the low-frequency power will steeper, thus
UBR will not prominent for flute. A slightly different pitch was essential
for UBR. For example, the keyboard instruments have fixed pitch and
the player cannot control the frequency thus UBR cannot be expected
for them. Therefore there may be no ensemble of keyboard instruments. 
\item In quantum mechanics, a typical double slit experiment can yield large-scale
interference or beat pattern for the propagating electron wave function.
In the same way, \textbf{the spatial distribution} of the sound sources
may also control the feature of UBR although we have not included
this effect in our present paper. The spatial arrangement of various
instruments in the orchestra, as well as the location of the audience,
would be essential for the variety of UBR.  
\item We made our sound samples by using Wolfram \textbf{Mathematica12}.
It would be interesting to use an electronic \textbf{synthesizer}
for the analysis of UBR because the synthesizer may produce more realistic
sound well mimicking the real musical instruments. We did not use
it this time simply because we do not know artificial filtering used
in typical synthesizers. 
\item The indicator UBR should be defined more elaborately. We have so far
characterized UBR by the power law and its power index. We probably
need the amplitude of the signal. For example, In Fig.\ref{melody},
we may define the indicator $R_{100-}^{0.1}$as the ratio of the PS
amplitudes at 0.1Hz and the maximum amplitude at beyond 100Hz. The
panels (a),(b),(c) respectively show $R_{100-}^{0.1}$ about $5\times10^{-5},20,500$.
This is reasonable since the overlap regions between the notes are
increasing in this order, $0\%,1\%,10\%$.
\item We have utilized numerical calculations in this paper, but a more
\textbf{analytic approach} will be possible. Although simple Fourier
transformation of the sound superposition only yields tremendous terms
of Dirac-delta functions, any sophisticated rearrangement of them
may make the data tractable. Further, the analytic approach is more
necessary in some situations in our analysis. For example, in the
resonance case, continuous eigenmodes contribute to the unison with
the weight described by the Cauchy\textendash Lorentz distribution
function. 
\item UBR we studied is characterized by the power-law of index $-1.5\thicksim-1$
in the PS, and the power-law seems to continue much lower frequency
domain. This feature reminds us of \textbf{1/f fluctuations} which
appear everywhere in nature. If our sound beat is accepted as a class
of \textbf{1/f fluctuations}, we would like to examine how extent
the \textbf{wave beats }can be a general mechanism of 1/f fluctuations.
We then further generalize UBR to the wave function of the electron
in semi-conductors. To proceed, we need to answer the objections raised
in \cite{Nieuwenhuizen1987,Kiss1986}. Probably, we need to extend
the ensemble to the stable distribution \cite{Cizek2005} characterized
by the property that a partial sum of the random fields, if scaled,
follows the same distribution. 
\item We have studied wave beat from the viewpoint of unison. This can also
be understood from the viewpoint of \textbf{synchronization} of many
active elements. For example, solar dynamo activity can be described
by the synchronization of many macro-spins each of which has a local
magnetic moment\cite{Nakamich2012}. This macro-spin model well describes
observed features of the solar activities. In particular, it describes
the 11year periodicity as well as 1/f fluctuations in the solar magnetism.
Actually, the PS of the sunspot record shows the power of index $-1.1$
on top of the 11year period\cite{Nakamich2012}. In this way, any
synchronizing system may show 1/f fluctuations by the interference
of elements.
\item We have studied wave beat within the time domain; high-frequency sound
sources synchronize with each other to yield the low-frequency structure
UBR. If further extended also to the spatial domain, we notice that
this mechanism can be understood as the transition from \textbf{the
microscopic to macroscopic} structures. \\
In this context, we can understand the generation of density fluctuations
in the early universe as the quantum beat of almost mass-less inflaton
field which also drove the inflationary cosmic expansion. This mechanism
will be deeply related to the quantum 1/f noise proposed in\cite{Handel1980}.
Actually, the present cosmic observations support the Zel'dovich spectrum
in the density fluctuations. This spectrum diverges both in ultraviolet
and infrared regions, the same as 1/f fluctuations in the time domain.
We need to clarify a possible common mechanism, if any, among primordial
density fluctuations, quantum 1/f noise, and UBR. 
\end{enumerate}
These extensions and applications are now under study in our group
and I hope we can report some of them soon. 
\begin{acknowledgments}
The author would like to thank many researchers and students of a
variety of expertise: members of the Astrophysics group, summer lecture
students 2021 organized by Mei Odo; Yayoi Abe, Katsuyoshi Kobayashi,
at Ochanomizu Univ.; Yutaka Shikano at Gunma Univ.; Koichiro Umetsu
and Toshiki Hanyu at Nihon Univ; and many others.
\end{acknowledgments}

\end{document}